\documentclass[prl,twocolumn,aps,superscriptaddress]{revtex4-1}
\usepackage{graphicx}
\usepackage{amsmath}
\usepackage[colorlinks=true,citecolor=blue, urlcolor=blue, linkcolor = blue]{hyperref}
\usepackage{amssymb}
\usepackage{cancel}
\usepackage{xcolor}
\usepackage{float} 
\usepackage[caption=false]{subfig}
\definecolor{lgray}{gray}{0.8}
\usepackage{verbatim} 
\usepackage{tabularx} 
\newcolumntype{Y}{>{\centering\arraybackslash}X} 
\usepackage{braket}
\usepackage{mathtools} 

\newcommand{\circled}[1]{\textcircled{\raisebox{-0.9pt}{#1}}}

\makeatletter
\DeclareRobustCommand{\iscircle}{\mathord{\mathpalette\is@circle\relax}}
\newcommand\is@circle[2]{%
	\begingroup
	\sbox\z@{\raisebox{\depth}{$\m@th#1\bigcirc$}}%
	\sbox\tw@{$#1\square$}%
	\resizebox{!}{\ht\tw@}{\usebox{\z@}}%
	\endgroup
}
\makeatother

\begin{document}

\title{Light-induced quantum droplet phases of lattice bosons in multimode cavities}

\author{P. Karpov}
\email{karpov@pks.mpg.de, karpov.petr@gmail.com}
\affiliation{Max Planck Institute for the Physics of Complex Systems, N{\"o}thnitzer Stra{\ss}e 38, Dresden 01187, Germany}
\affiliation{National University of Science and Technology ``MISiS'', Moscow, Russia}

\author{F. Piazza}
\email{piazza@pks.mpg.de}
\affiliation{Max Planck Institute for the Physics of Complex Systems, N{\"o}thnitzer Stra{\ss}e 38, Dresden 01187, Germany}

\date{June 24, 2021}

\begin{abstract}

Multimode optical cavities can be used to implement interatomic interactions which are highly tunable in strength and range.
For bosonic atoms trapped in an optical lattice, cavity-mediated interactions compete with the short-range interatomic repulsion, which we study using an extended Bose-Hubbard model. Already in a single-mode cavity, where the corresponding interaction has an infinite range, a rich phase diagram has been experimentally observed, featuring density-wave and supersolid self-organized phases in addition to the usual superfluid and Mott insulator. Here we show that, for any finite range of the cavity-mediated interaction, quantum self-bound droplets dominate the ground state phase diagram. Their size and in turn density is not externally fixed but rather emerges from the competition between local repulsion and finite-range attraction. Therefore, the phase diagram becomes very rich, featuring both compressible superfluid/supersolid as well as incompressible Mott and density-wave droplets. Additionally, we observe droplets with a compressible core and incompressible outer shells.

\end{abstract}

\maketitle


\emph{Introduction.} 
Ultracold atoms in optical cavities represent a unique experimental platform for the study of strongly interacting quantum many-body systems of light and matter \cite{Ritsch:2013,Piazza:2020}.
The cavity-mediated interactions are naturally long-ranged and can easily overcome the short-range interatomic repulsion. The competition between these two types of interaction can be enhanced by trapping the atoms in an optical lattice. Already for a single-mode cavity, which mediates global-range interactions, a rich phase diagram has been experimentally observed, featuring, besides the known superfluid and Mott-insulating phases, a lattice supersolid as well as an incompressible density-wave phase \cite{Landig:2016}. Such experiments implement a Bose-Hubbard (BH) model extended by a global-range sign-changing interaction, which has been intensively studied in recent years 
\cite{Li:2013,Bakhtiari:2015,Dogra:2016,Chen:2016,Sundar:2016,Flottat:2017,Panas:2017,Chiacchio:2018,Nagy:2018,Liao:2018,Himbert:2019}.

More recently, the coupling to multiple modes of the cavity has emerged as a very promising route towards an even richer many-body physics. Already the use of two cavity modes has allowed to observe a supersolid with continuous translational symmetry breaking \cite{Leonard:2017,Lang:2017,Mivehvar:2018,Schuster:2020}.
By tuning cavitites around the confocal degenerate point \cite{Kollar:2015,Kollar:2017}, it has recently become possible to even control the range of the cavity-mediated interaction \cite{Vaidya:2018,Guo:2019}, and to realize an optical lattice featuring phonons \cite{Guo:2021}.
With finite-range cavity-mediated interactions, supersolids have been predicted to feature
crystalline topological defects and to appear through non-mean-field phase transition dominated by fluctuations \cite{Gopalakrishnan:2009, Gopalakrishnan:2010}.


Here we demonstrate that, due to the competition between cavity-mediated finite-range attraction and the on-site repulsion, the phase diagram is actually dominated by a variety of droplet phases, i.e. self-bound many-particle quantum objects, while the extended phases exist only for sufficient repulsion (see \cite{Karpov:2019} for the classical, purely attractive case).  
We study a BH model extended by an attractive tunable-range interaction, with and without additional sing-changing modulation, which can be realized in state-of-the-art confocal-cavity experiments. We compute the ground-state phase diagram with a worm quantum Monte Carlo algorithm, using a canonical version  \cite{Rombouts:2006,Houcke:2006,Pollet:2005} necessary to describe droplet phases.
We observe a complex sequence of transitions between droplets of different sizes, and of compressible (superfluid or supersolid) as well as incompressible (Mott or density-wave insulating) nature, governed by the competition between the local repulsion and the finite-range attraction. Within the superfluid-droplet phase, we additionally observe a sequence of crossovers between the fully superfluid droplets and droplets with superfluid core and Mott-like outer shells.


Without the lattice, quantum droplets of bosons have been experimentally observed using dipolar interactions \cite{Pfau:2016-Nature,Chomaz:2016} and bosonic mixtures \cite{Tarruell:2018-Science,Semeghini:2018}.
In dipolar systems the anisotropic interactions can lead quantum-coherent (supersolid) chains of droplets that are metastable due to three-body losses \cite{Pollet:2019, Tanzi:2019, Bottcher:2019, Chomaz:2019}, as well as other droplet arrangements and non-trivial density patterns \cite{Boninsegni:2019, Hertkorn:2021, Zhang:2021}.

Lattice bosons with dipolar interactions \cite{Goral:2002, Baier:2016} and mixtures \cite{Morera:2020, Morera:2021} have been proposed for implementing various types of extended Hubbard models \cite{Dutta:2015}.
While currently for dipolar gases and bosonic mixtures the challenge is to reach appreciable interactions beyond the nearest neighbour, cavity-mediated interactions are instead naturally strong at large distances. This feature is shared by other types of photon-mediated interactions based on refraction \cite{Ostermann:2016} or diffraction \cite{Labeyrie:2014,Tesio:2014,Robb:2015} (the latter also predicted to induce droplet phases \cite{Pohl:2018,Pohl:2021}), which however lack the tunability of the range.

\begin{figure}[!t] 
	\centering
	\includegraphics[width=1.0\linewidth]{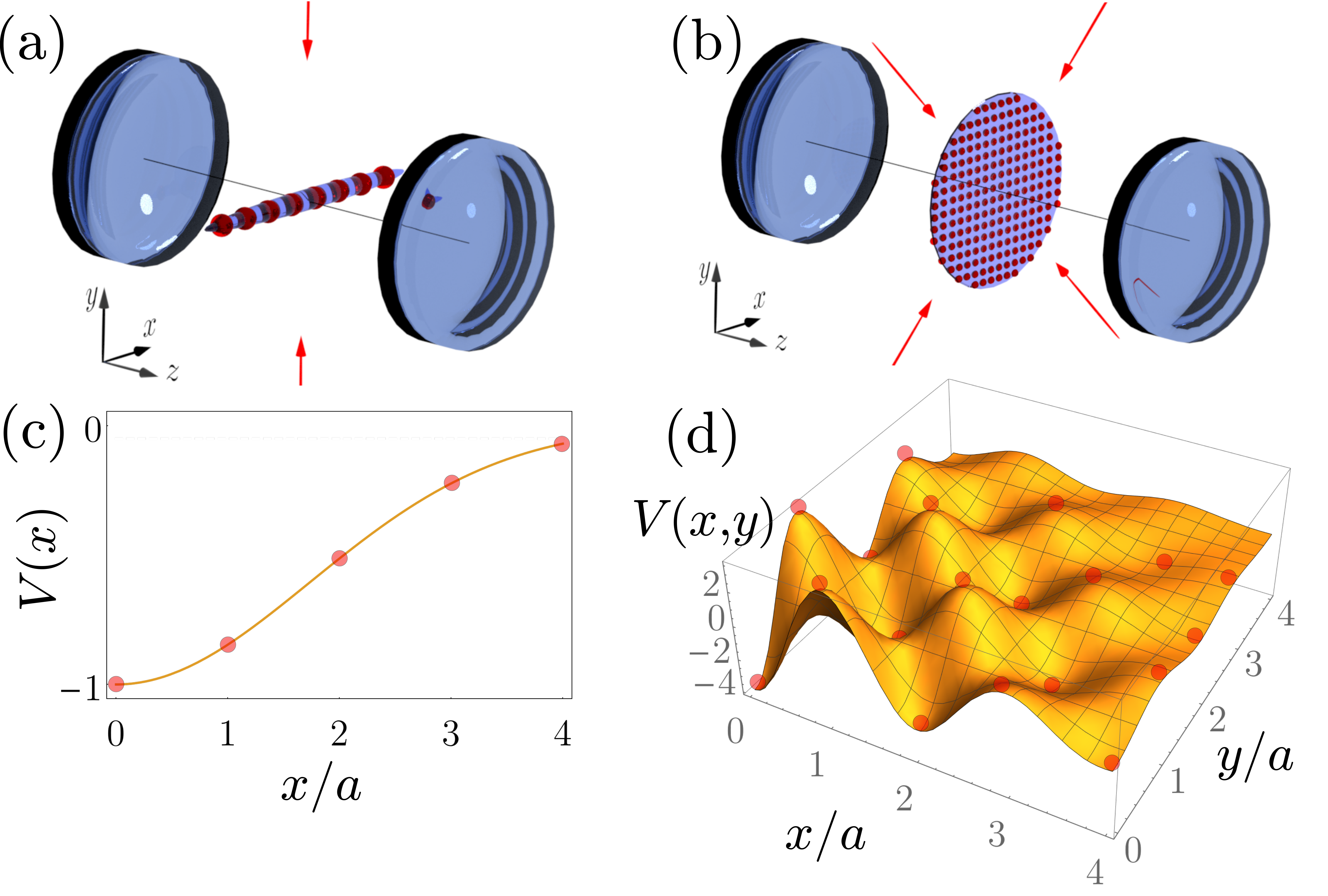}
	\caption{
          We consider ultracold bosons in two configurations corresponding to two different cavity-mediated interaction potentials.
          The latter are generated in the dispersive regime via two-photon transitions involving retroreflected lasers beams (red arrows) which are red detuned from a given degenerate family of a multimode cavity. The interaction potential results from the interference between the laser and the cavity field.
            In (a), a quasi 1D gas is trapped in an optical lattice within the midplane of the cavity: $z=0$, along the $x$ direction perpendicular to the laser. This results into the monotonous interaction depicted in (c), which decays over a scale $\xi$ controlled by the number of degenerate, transverse cavity modes \cite{Vaidya:2018,Guo:2019A}. In (b), the lattice gas is quasi-2D and two perpendicular retroreflected laser modes interfere with the cavity field. This induces the sign-changing interaction depicted in (d). The trapping optical-lattice potentials are generated by three additional far-off detuned lasers (oriented along the $x$, $y$, and $z$ directions, not shown) which do not interfere with the cavity field \cite{Landig:2016}.}        
	\label{fig:geometry}
\end{figure}

\begin{figure*}[!hbt] 
	\centering
	\includegraphics[width=1.0\linewidth]{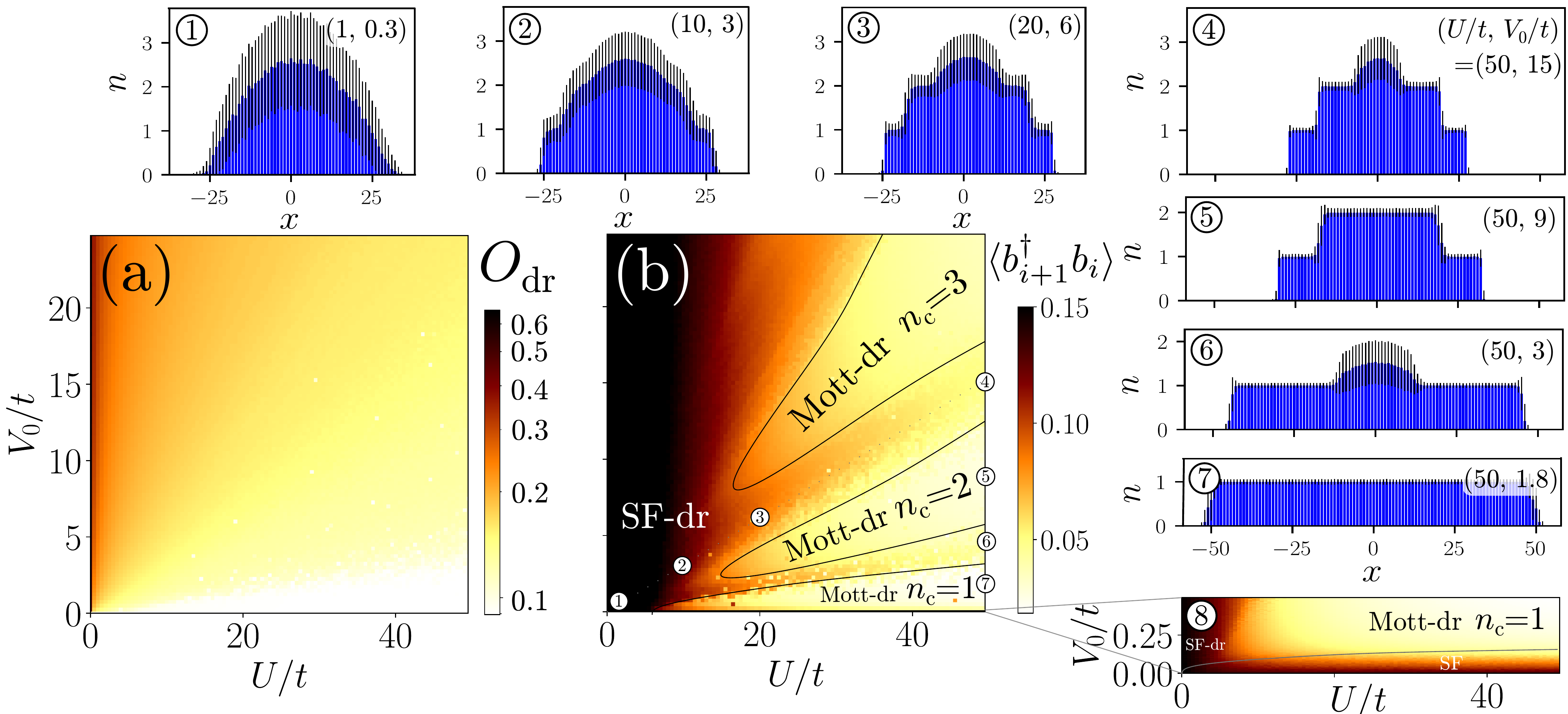}
	\caption{
		The ground-state phase diagram of the 1D system with monotonous interactions in the coordinates $V_0$ (long-range attraction) vs $U$ (on-site repulsion). The droplet thermodynamic limit $\xi, N\rightarrow\infty$ with $n_0=N/\xi=0.8$ is calculated by extrapolation up to $L=500$, $N=100$, $\xi=125$.
		(a) Droplet order parameter  $O_{\mathrm{dr}}$.
		(b) Nearest-neighbor superfluid correlator $\langle b^{\dagger}_{i+1} b_i \rangle$. 
		Insets \circled{1}-\circled{7} show the spatial density profiles of the droplets at the corresponding points in the phase diagram; the error bars show the standard deviation of density from its mean value.
		Inset \circled{8} shows the SF-correlator for in the region $V_0/t<0.5$ featuring the SF phase. 
	}
	\label{fig:phase-diagram_monotonous}
\end{figure*}

\emph{Model.}
We consider ultracold bosons trapped in an optical lattice in the two different configurations illustrated in Fig.~\ref{fig:geometry}. A quasi-1D geometry with monotonous cavity-mediated interactions, and a quasi-2D with sign-changing interactions. As derived in detail in the Supplementary Material, the system can be described by an extended BH model 
\begin{align}
H = -t \sum_{\langle \mathbf{i}, \,\mathbf{j}\rangle} (b_\mathbf{i}^{\dagger} b_\mathbf{j} + h.c.) + \frac{U}{2}  \sum_{\mathbf{i}} n_\mathbf{i} (n_\mathbf{i}-1) + \sum_{ \mathbf{i}, \,\mathbf{j}} V_{\mathbf{i},\,\mathbf{j}} n_\mathbf{i} n_\mathbf{j}
\label{eq:H_Hubbard_U_V}
\end{align}
Here $\langle \mathbf{i}, \mathbf{j}\rangle$ denotes nearest-neighbor sites in a $D$ dimensional square lattice, $t$ is the hopping between the nearest neighboring sites of the optical lattice, $b_\mathbf{i}^{\dagger}$ and $b_\mathbf{i}$ are bosonic creation and annihilation operators, $n_\mathbf{i}=b_\mathbf{i}^{\dagger} b_\mathbf{i}$, $U>0$ is the on-site contact repulsion, and $V_{\mathbf{i},\,\mathbf{j}}$ is the tunable-range interaction mediated by the cavity between particles at sites $\mathbf{i}$ and $\mathbf{j}$. 
We consider a multimode optical cavity, 
where the finite range $\xi$ is achieved by having a large number of transverse modes within a given degenerate family. Alternatively, Floquet modulation \cite{Johansen:2020} or multifrequency driving \cite{Torggler:2014} can render the range finite even for non-degenerate cavities. Experiments with ultracold bosons in confocal cavities have demonstrated the ability to tune this range \cite{Vaidya:2018}.
We model this using the following generic form of the interaction potential: 
\begin{align}
V_{\mathbf{i},\,\mathbf{j}}\equiv V_{\mathbf{i}-\mathbf{j}} = 
-s_{\mathbf{i}-\mathbf{j}}  V_0 \exp\left(-\frac{|\mathbf{i}-\mathbf{j}|^2}{\xi^2}\right),
\label{eq:V}
\end{align}
where $V_0>0$ (locally attractive interaction). The sign-factor $s$ is either $s_{\mathbf{i}}=1$ for the monotonous attractive interaction or $s_{\mathbf{i}}=(-1)^{(i_x+i_y)}$ for the sign-changing interaction. All lengths are measured in the unit of lattice spacing. We stress that the results presented in the following do not qualitatively depend on the particular form of the envelope decay, as long as a single scale $\xi$ can be associated with it.

The length scale $\xi$ introduced above determines the characteristic radius of the droplet and helps to define a proper thermodynamic limit for a droplet phase, which is given by $L\rightarrow\infty$, $N\rightarrow\infty$, $\xi\rightarrow\infty$, $V_0 \rightarrow 0$, while $V_0\xi^D=\mathrm{const}$ and $N/\xi^D = \mathrm{const} \simeq n_{0}$. Here $V_0\xi^D$ gives the characteristic integrated interaction strength and the ``droplet filling'' $n_0$ gives the characteristic density of particles in a droplet of radius $\xi$.
A further important parameter is the characteristic slope of the density profile, $\sigma \simeq dn/dr \simeq n_{0}/\xi = N/\xi^{D+1}$ (see Supplementary Material). This steepness parameter vanishes in the droplet thermodynamic limit defined above, and quantifies the amount of finite-size effects in the phase diagram. As we discuss later and in the Supplementary Material, these finite size effects are experimentally measurable and it is important to understand them for the correct extrapolation to the thermodynamic limit.

\emph{Method.}
We map the ground-state phase diagram of the model using a canonical worm Quantum Monte Carlo algorithm \cite{Rombouts:2006,Houcke:2006,Pollet:2005} with fixed total number of particles $N=\sum_\mathbf{i} n_\mathbf{i}$  and periodic boundary conditions.
Since we explicitly work in the canonical ensemble, we can deal with arbitrary occupation numbers (up to the total number of particles in the system) and thus for the search of potential droplet phases the method is superior with respect to other traditional methods of studying the BH model, like grand-canonical quantum Monte Carlo \cite{Prokofiev:1998, Sandvik:1999, Syljuaasen:2002} methods or DMRG \cite{Schollwock:2005}, which require fine tuning of the chemical potential and/or cut-offs for the local occupation.
In order to construct the ground-state phase diagram of the model, we perform calculations at small non-zero temperature $T=0.1 t$ (and check that this temperature is low enough by comparing our method with Lanczos exact diagonalization up to $L=N=16$ systems in 1D).

We characterize the droplets using two observables: the droplet order parameter $O_{\mathrm{dr}} = \sqrt{\frac{N_s}{N_s-1} \sum_i \left(\frac{n_i}{N}-\frac{1}{N_s} \right)^2}$ where $N_s=L^D$, and the single-particle superfluid correlations between neighboring sites $\langle b^{\dagger}_{i+1} b_i\rangle$ (in 1D monotonous interaction case) and next-neighboring sites $\langle b^{\dagger}_{i_x+1,i_y+1} b_{i_x,i_y}\rangle$ (in 2D sign-changing interaction case), averaged over all sites $\mathbf{i}$.

\emph{Monotonous interaction.} 
Figure \ref{fig:phase-diagram_monotonous} shows the ground-state phase diagram for the case of monotonous interaction $s_{i,j}=1$ in 1D (setting of Fig.~\ref{fig:geometry}(a)). 

For $U=V_0=0$ the system is in the SF state. With increasing $V_0$ we enter a single-site-droplet phase at  $V_0/t \sim 1/N$ \cite{Kanamoto:2003}. Upon increasing $U$ and decreasing $V_0$ we observe smooth crossovers between droplets of different ``diameters'' (Fig.~\ref{fig:phase-diagram_monotonous}(a)), up to the maximum diameter bounded either by the size of the system or the total number of particles $d_{\mathrm{max}}\simeq \min(L/2, N)$.
Droplets whose diameter is larger than $\xi$ are only weakly bound as the outer particles are not bound to the center.
After the maximum droplet size is reached, a first-order transition to a phase with uniform density occurs, either a Mott insulator for integer fillings or a SF for non-integer fillings.

Fig.~\ref{fig:phase-diagram_monotonous}(b), shows the nearest-neighbor superfluid correlators, where we observe the various superfluid and Mott quantum droplet phases.
The Mott phases here correspond to different fillings at the core of the droplet, e.g. $n_c=1,2,3$. 

The existence of the transitions between SF and Mott droplets can be qualitatively understood by means of the standard Bose-Hubbard phase diagram \cite{Fisher:1989}, whereby the chemical potential is set by the local occupation at the droplet core.
For $V_0=0$ and a fixed integer average filling $\nu=N/L$, upon increasing $U$ we have a conventional SF-Mott transition at $U/t\approx4\nu$ \cite{Oosten:2001}.
Analogously, for the droplet phases, with increasing $U$ along lines $V_0/U=\mathrm{const}$ the superfluid correlator between neighboring sites shows SF-Mott transitions.
We emphasize however that, unlike the case of the Mott-SF transitions in an external confining potential, here the droplet size and density distribution is not externally fixed but rather emerges from the competition between local repulsion and finite-range attraction. Moreover, the droplet spontaneously breaks the translation symmetry of the Hamiltonian.

Additionally, we compute the real-space density profiles of the droplets (insets \circled{1}-\circled{7} of Fig.~\ref{fig:phase-diagram_monotonous}(b)).
Moving outwards in the radial direction of the phase diagram we first observe the fully superfluid droplet \circled{1} (see Supplementary Materials for an analytic Thomas-Fermi description), which gradually develops Mott-type outer shells, and finally crosses over to a droplet with only the core being superfluid \circled{4}.
Moving along the vertical line $U/t=50$ we observe instead transitions between SF-droplets \circled{4}, \circled{6} and Mott droplets \circled{5}, \circled{7}.  

The inset \circled{8} of Fig.~\ref{fig:phase-diagram_monotonous}(b) shows a zoom in the $V_0/t<0.5$ part of the phase diagram, featuring the extended SF phase. Using (\ref{eq:H_Hubbard_U_V}) we can estimate the critical value of $V_0$ for the transition between the Mott droplet with $n_c=1$ and the SF-phase. In the former phase the energy is dominated by the interaction part $E_{\mathrm{Mott-dr}}\sim -N V_0 \xi^D$, while in the latter phase the energy is dominated by the kinetic part $E_{\mathrm{SF}}\sim -N t$. Equating the two energies, the transition value can be estimated as $V_0\sim t/\xi^D$.

We have explicitly checked (finite-size scaling data from $L=100$, $N=20$, $\xi=25$ to $L=500$, $N=100$, $\xi=125$ is not shown) that the droplet regime of the phase diagram has a well-defined thermodynamic limit, $L, N, \xi\rightarrow\infty$, while $N/\xi = \mathrm{const}$. For finite-size droplets featuring a steepness parameter $\sigma \gtrsim 1$, we observe additional features, e.g. mesoscopic first-order phase transitions associated with change in the droplet size (see Supplementary Material). Similar examples are visible in the 2D calculations shown next.

\begin{figure}[!t] 
	\centering
	\includegraphics[width=1.0\linewidth]{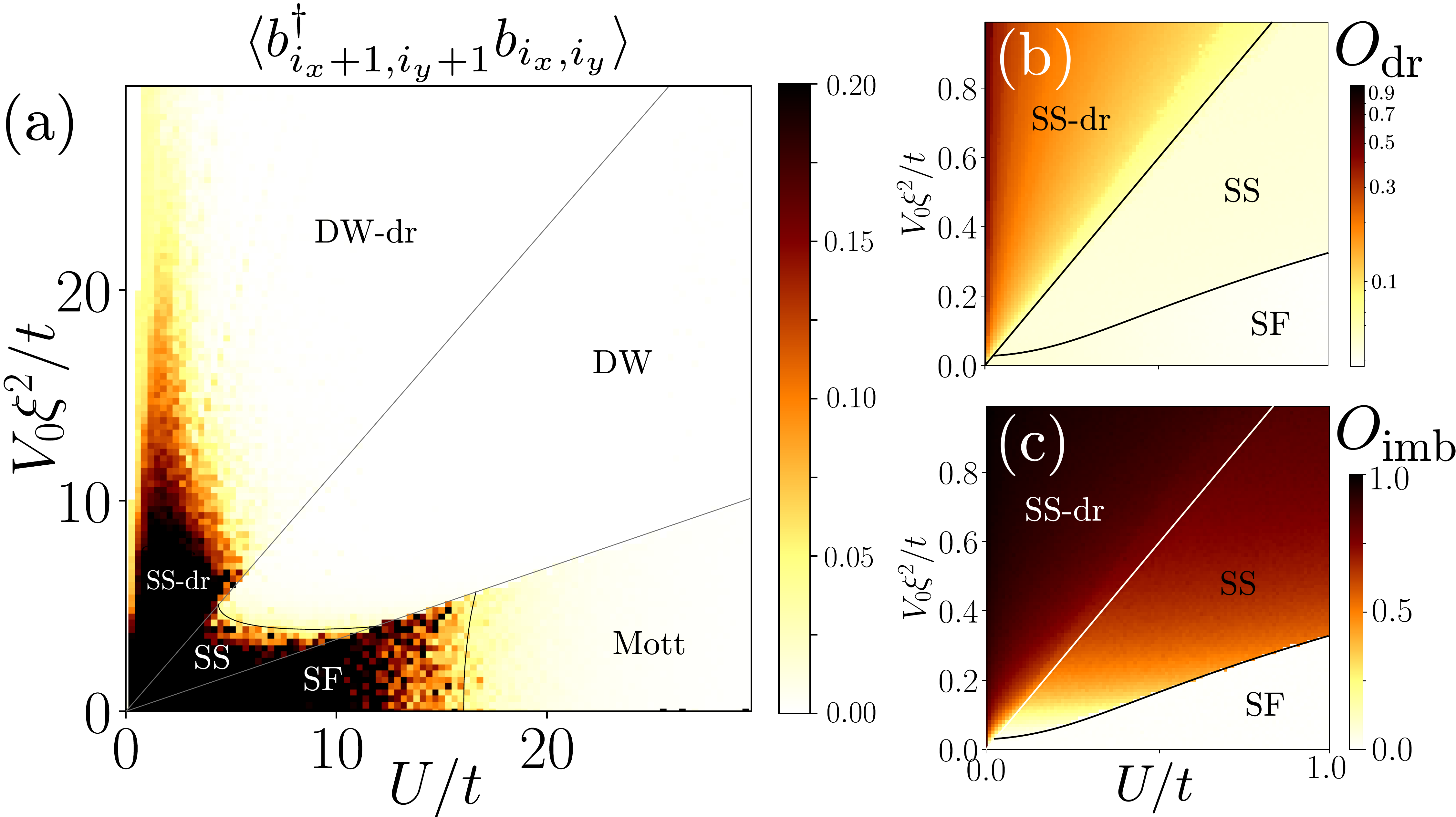}
	\caption{
		Phase diagram for the 2D system with sign-changing interaction
		for unit lattice filling $N/L^2=1$  and $n_0=N/\xi^2=16$, $\sigma=N/\xi^3=3.2$ ($L=20$, $N=400$, $\xi=5$).
		(a) Superfluid correlator  $\langle b^{\dagger}_{i_x+1,i_y+1} b_{i_x,i_y} \rangle$.
		(b) Droplet order parameter  $O_{\mathrm{dr}}$ and
		(c) sublattice imbalance order parameter $O_{\mathrm{imb}}=(N_{\mathrm{even}}-N_{\mathrm{odd}})/(N_{\mathrm{even}}+N_{\mathrm{odd}})$ zoomed in the vicinity of the origin, $U/t$, $V_0\xi^2/t < 1$.
	}
	\label{fig:phase-diagram_sign-changing_xi=1}
\end{figure}
\emph{Sign-changing interaction.}
Now we turn to the case of the sign-changing interaction, which makes possible the existence of the density-wave (DW)  and supersolid (SS) phases. We consider the two-dimensional setting of Fig.~\ref{fig:geometry}(b).

The infinite-range case, $\xi=\infty$, has been studied experimentally in a single-mode cavity \cite{Landig:2016}. The phase diagram we find in the case of a finite range, shown in Fig.~\ref{fig:phase-diagram_sign-changing_xi=1} for the relatively short range $\xi=1$ and unit filling $n=N/L^2=1$, features the same extended phases as the infinite-range case (studied with QMC in \cite{Flottat:2017}), plus quantum droplet phases, which occupy a substantial part of the phase diagram.
Note that in order to perform the comparison between the $\xi=\infty$ and finite $\xi$ cases we use the rescaled coordinate of vertical axis, $V_0 \xi^D/t$.

For $V_0=0$ the system is either in the SF phase (at $U/t\lesssim15$) or the Mott phase  (at $U/t\gtrsim15$, Fig.~\ref{fig:phase-diagram_sign-changing_xi=1}). Increasing $V_0$ favors the density-modulated SS and DW phases.
In these phases we can estimate an effective hopping within a given sublattice using second-order perturbation theory.
In extended SS and DW phases, as well as for droplets with $d\gg\xi$, the effective hopping between sites of the even (odd) checkerboard sublattice is $t_{\mathrm{eff}} \sim n_0 t^2/(\gamma n_0 V_0 \xi^D - (n_0-1) U)$, with $\gamma$ some numerical constant. 
For smaller droplets with $d\lesssim\xi$ the corresponding estimate gives $t_{\mathrm{eff}} \sim  n_0 t^2/(\gamma N V_0  - (n_0-1)U)$.
In both cases, upon moving in the radial direction in the phase diagram the extended SS and the SS-droplet phases feature much smaller SF correlators in comparison to the case of monotonous interactions, where $\braket{b_{i+1}^{\dagger} b_i} \sim t/U$.
This is due to the fact that for the density-modulated phases $\langle b^{\dagger}_{i_x+1,i_y+1} b_{i_x,i_y} \rangle \sim t_{\mathrm{eff}}/U \sim t^2/U^2$ acquires an additional $t/U$ suppression factor for increasing $U,V_0 \xi^D$ and $U/V_0\xi^D=\mathrm{const}$.
Hence, the lobe structure for different DW-droplet phases is not visible in Fig.~\ref{fig:phase-diagram_sign-changing_xi=1}(a). Still, $t_{\mathrm{eff}}$ is finite in the thermodynamic limit $N,\xi\rightarrow\infty$, $V_0\rightarrow0$, since $n_0=N/\xi^D$, $V_0\xi^D$, as well as their product, $N V_0$, are constant.

\begin{figure}[!t] 
	\centering
	\includegraphics[width=1.0\linewidth]{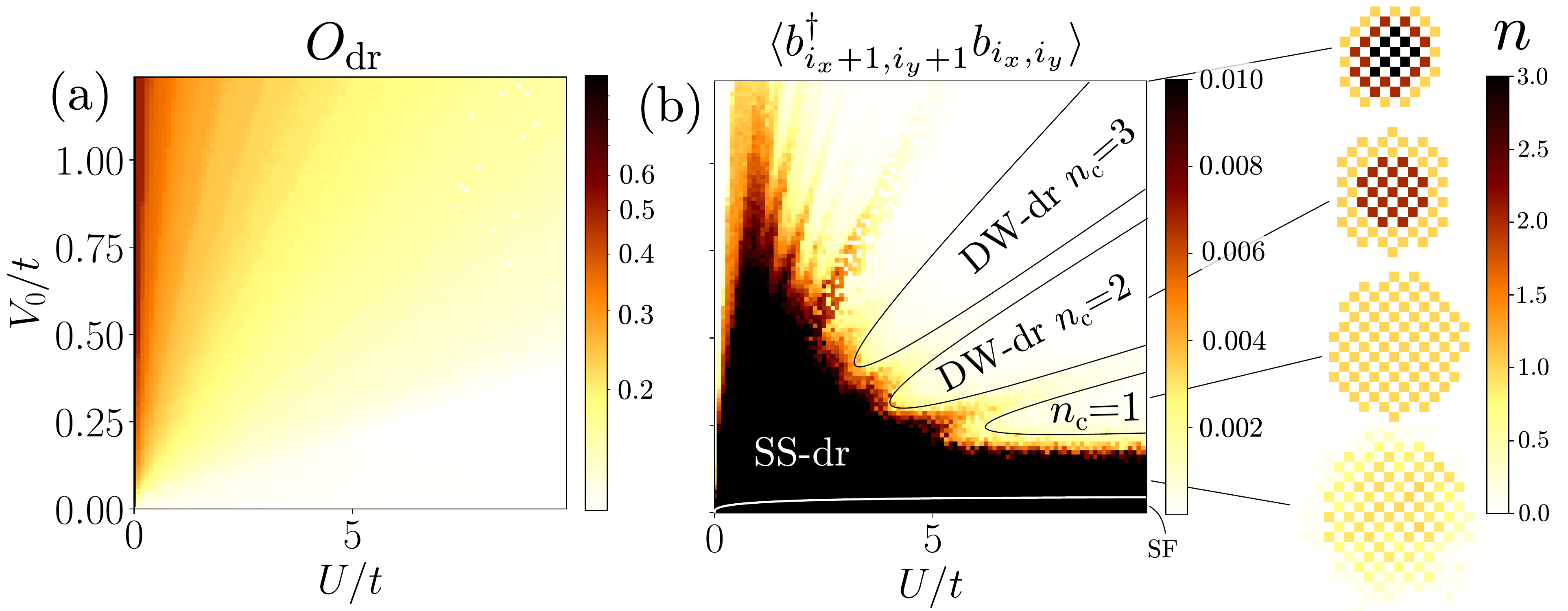}
	\caption{
		Phase diagram for 2D system with sign-changing interactions for small lattice filling $N/L^2\approx0.08$, and $n_0=N/\xi^2=1.25$  $\sigma=N/\xi^3\approx0.16$ ($L=32$, $N=80$, $\xi=8$).
		(a) Droplet order parameter  $O_{\mathrm{dr}}$. 
		(b) Superfluid correlator $\langle b^{\dagger}_{i_x+1,i_y+1} b_{i_x,i_y} \rangle$; in order to emphasize the Mott region we cut off the color scale beyond values of the correlator $>0.01$; the insets show the droplet density distribution at the corresponding points of phase diagram.
	}
	\label{fig:phase-diagram_sign-changing_flat_droplet}
\end{figure}

In order to make the lobe structure of the quantum droplet phases visible, in Fig.~\ref{fig:phase-diagram_sign-changing_flat_droplet} we show the phase diagram for a small lattice filling 
 $n\approx0.08$ and low steepness parameter $\sigma \approx 0.16$ (see Supplementary Material for the finite-size case with large $\sigma$).
Like for the monotonous interaction, we observe a non-trivial change in the superfluid correlator in the azimuthal direction (Fig.~\ref{fig:phase-diagram_sign-changing_flat_droplet}(b)), due to appearance of the self-consistent DW-droplet lobes separated by regions of superfluid droplets. 
The insets of Fig.~\ref{fig:phase-diagram_sign-changing_flat_droplet}(c) show typical Mott-droplet real-space density profiles.
We note that here $\sigma$ is not small enough to eliminate finite-size effects: in the upper-left corner of Fig.~\ref{fig:phase-diagram_sign-changing_flat_droplet}(a) we see a sequence of first-order transitions between superfluid droplets of different sizes.

\emph{Conclusions and outlook.}
In this work, we have proposed an experimental setting for the creation self-bound quantum droplets for bosonic particles in an optical lattice, where the finite-range attractive tunable interaction is induced by a multimode optical cavity and competes with the local repulsion. We modelled such a system by a rather generic class of extended Bose-Hubbard models. We observed various types of lattice droplet phases: superfluid/Mott droplets for the monotonous finite-range interactions and supersolid/density-wave droplets for sign-changing interactions.
Unlike the case of lattice bosons in a trap, droplets are self-bound objects which break the translation symmetry spontaneously, so that the local chemical potential emerges from the competition between local repulsion and finite-range attraction.

Other unexplored classes of extended BH-models related to the one studied here might be also realized by trapping dipolar Bose gases \cite{Pfau:2016-Nature,Chomaz:2016,Pollet:2019, Tanzi:2019, Bottcher:2019, Chomaz:2019} or bosonic mixtures \cite{Tarruell:2018-Science,Semeghini:2018} in a lattice. The numerical methods and the analysis employed here should offer a helpful guide for the future investigation of new quantum droplet phases in such models.

Under realistic conditions for large local densities of particles also three-body effects, not considered here, can play an important role and show up either as a \emph{three-body interactions} \cite{Bulgac:2002, Bulgac:2003, Dutta:2015}, favoring extended droplets, or \emph{three-body losses}, making the system metastable, like in the recent dipolar gas experiments \cite{Tanzi:2019, Bottcher:2019, Chomaz:2019}.

Finally, we mention that the droplet physics discussed here is rather generic and our analysis might also be relevant in other contexts, for example for the quantum simulation of black holes as real-space Bose-Einstein condensates of gravitons \cite{Dvali:2014}.

\emph{Acknowledgments.}
We thank Tobias Donner, Panos Giannakeas, David Luitz, Alessio Recati, and Maurits Tepaske for useful discussions. 
P. K. acknowledges the support of the Alexander von Humboldt Foundation and the Ministry of Science and
Higher Education of the Russian Federation.

\bibliographystyle{apsrev4-1}
\bibliography{literature_quantum_droplets}



\clearpage
\widetext
\begin{center}
	\textbf{\large Supplemental Materials: Light-induced quantum droplet phases of lattice bosons in multimode cavities}
\end{center}
\setcounter{equation}{0}
\setcounter{figure}{0}
\setcounter{page}{1}
\makeatletter
\renewcommand{\theequation}{S\arabic{equation}}
\renewcommand{\thefigure}{S\arabic{figure}}

\setcounter{secnumdepth}{3}  

\section{Thomas-Fermi droplet-profile and in the continuum limit}

\begin{figure}[!tbh] 
	\centering
	\includegraphics[width=1.0\linewidth]{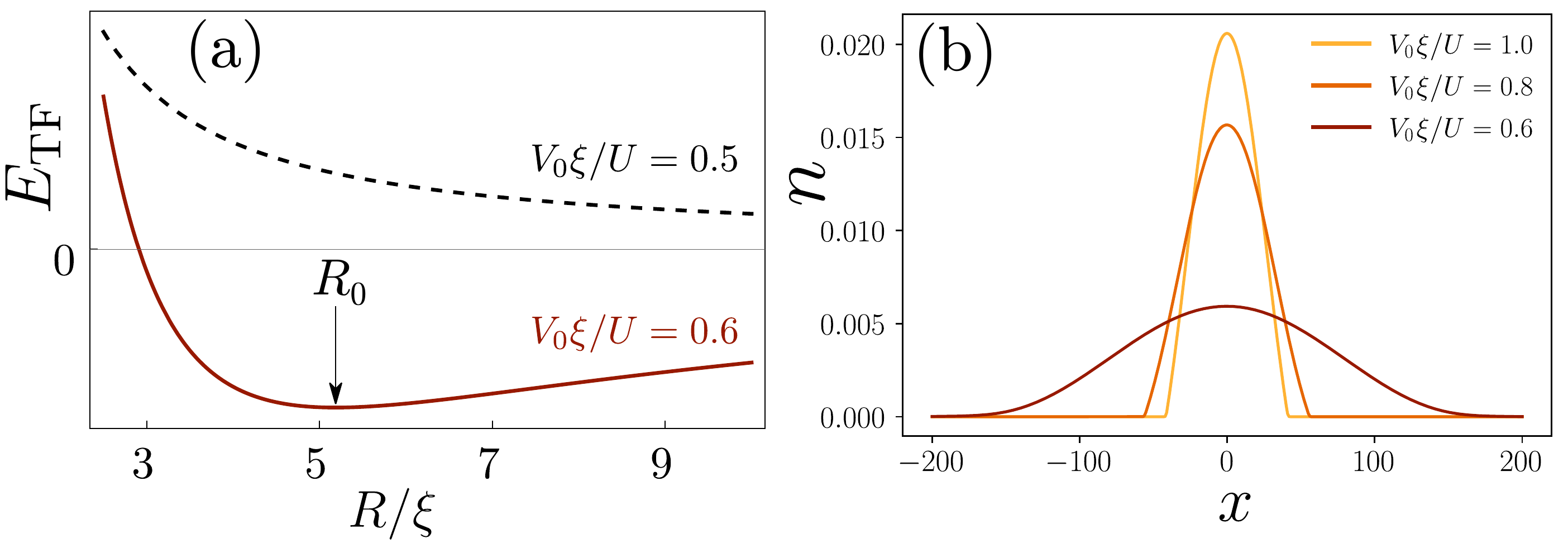}
	\caption{
		The Thomas-Fermi picture for the monotonous interaction in 1D.
		(a) The Thomas-Fermi energy $E_{\mathrm{TF}}$ (in arbitrary units) versus the radius of the droplet $R/\xi$, given by eqs. (\ref{suppl_eq:E_TF})-(\ref{suppl_eq:f_rho}) for $V_0\xi/U=0.6$ (droplet state with radius $R_0$) and $0.5$ (uniform state). For the actual system with discrete fillings the latter case corresponds to the droplet with local density $=1$ particle per site (average density $<1$ is assumed).
		(b) The Thomas-Fermi density profiles $n(x)$ for the  for different values of $V_0\xi/U$. 
		Here we use $\xi=25$, the density is calculated by the imaginary time propagation of GPE with small hopping $t=0.01 V_0$, and normalization $\int n(x)dx = 1$.
	}
	\label{suppl_fig:Thomas-Fermi}
\end{figure}

In this Supplementary Section we perform the Thomas-Fermi analysis of the superfluid droplets and find their density profiles.

Here we will work in the continuum limit: i.e. the limit when all characteristic length scales of the problem such as the range of the interaction potential and the radius of the droplet $\xi,R\gg 1$ (let us recall that all lengths are measured in the units of lattice constant $a$), and the local occupation numbers $n\gg1$ in the core of the the droplet. For simplicity, we will work in one dimension.

The density profile $ n_{\mathrm{TF}}  \equiv n(x)$ in the Thomas-Fermi approximation in a self-consistent form is given by
\begin{align}
n(x) = \frac{1}{U} (\mu-V_{\mathrm{eff}}(x)) \theta(\mu-V_{\mathrm{eff}}(x)) 
\end{align}
where $V_{\mathrm{eff}}(x) = \int_{-R}^{R} n(x') V(x-x') dx'$ is the effective potential experienced by a particle placed at $x$, $V(x)=-V_0 \exp(-x^2/\xi^2)$ is the interaction potential, and $\theta$ is the Heaviside step function.
Instead of solving the above integral equation we choose an ansatz for the density profile 
\begin{align}
n(x) = \frac{3}{4R} \left(1-\frac{x^2}{R^2}\right)
\end{align}
and optimize it with respect to the parameter $R$ (the radius of the droplet). Normalization is chosen in such a way that $\int_{-R}^{R} n(x) dx =1$. Note that we can systematically improve the ansatz by introducing the higher even powers of $x$ and the corresponding variational parameters.

The total Thomas-Fermi energy of the system can be written in the continuous limit as 
\begin{align}
E_{\mathrm{TF}} = E_U + E_V
\label{suppl_eq:E_TF}
\end{align}
where
\begin{align}
E_U = \frac{U}{2} \sum_i n_i^2 \rightarrow \frac{U}{2} \int_{-R}^{R} n^2(x) dx = \frac{3 U}{10 R},
\end{align}
\begin{align}
E_V =\frac{1}{2} \sum_{i,j} V_{i,j} n_i n_j \rightarrow \frac{1}{2} \int_{-R}^{R} V(x-x') n(x) n(x') dx dx'= -\frac{3V_0}{160} f(R/\xi),
\end{align}
and
\begin{align}
f(\rho)=  \frac{1}{\rho^6} \left( 2\sqrt{\pi } \rho^3 (8 \rho^2-5) \text{erf} (2\rho) +  e^{-4\rho^2} ( 1+8 \rho^4 - 6  \rho^2) +10 \rho^2 - 1 \right)
\label{suppl_eq:f_rho}
\end{align}

Let us denote the relevant droplet radius minimizing the Thomas-Fermi energy by $R_0$ and consider two limiting cases: $R_0/\xi \ll 1$ and $R_0/\xi \gg 1$.

a)  $R_0/\xi \ll 1$:
\begin{align}
E_{\mathrm{TF}}(R) = \frac{3U}{10R} + \frac{V_0 R^2}{5 \xi^2} + \mathrm{const}
\end{align}
Minimizing $E_{\mathrm{TF}}(R)$ with respect to $R$ we get
\begin{align}
R_0 = \left(\frac{3U \xi^2}{4V_0}\right)^{1/3}
\label{suppl_eq:R0}
\end{align}
The condition $R_0/\xi \ll 1$ is equivalent to $V_0\xi\gg U$, which limits the applicability of the above formula. Extrapolating it to the values $V_0\xi \sim U$ (i.e. beyond its direct limit of applicability) we get an estimate $R_0^{\mathrm{max}} \sim \xi$.

b) $R_0/\xi \gg 1$ ($V_0\xi\ll U$):
\begin{align}
E_{\mathrm{TF}}(R) = \frac{3(U-\sqrt{\pi} V_0\xi)}{10R}
\end{align}
Here $E_{\mathrm{TF}}(R)$ is a monotonously decreasing function, so $R_0 \rightarrow \infty$. In the continuum approximation this means that the droplet increases till $R_0=L$ and the system goes to a uniform state. 
However, for the actual system with discrete fillings even the limit $U/(V_0\xi) \rightarrow \infty$ corresponds to the droplet with local density $=1$ particle per site (average density $<1$ is assumed). In this case, the Thomas-Fermi approximation breaks down and we have exactly $E_U=0$; the size of the droplet is limited by the total number of particles: $2R \simeq N$.

The critical value that separates these two regimes of the localized droplet and of the uniform solution can be found numerically and is equal to $(V_0\xi/U)_c \approx 0.565$. Figure \ref{suppl_fig:Thomas-Fermi}(a) shows the $E_{\mathrm{TF}}(R)$  dependence in both regimes.

Now let us analyze the droplet-size-changing and SF-Mott phase transitions.
In order to do this let us restore for a moment the discrete formulation of the problem with $N$ particles on a lattice.

The transitions between different droplet configurations (observed as sharp changes in the droplet order parameter $O_{\mathrm{dr}}$, see e.g. Fig.~\ref{fig:phase-diagram_monotonous}) correspond to changes of the radius by $\Delta R \sim1$ lattice site:
\begin{align}
\Delta V^{\mathrm{dr}}_0 = \frac{\partial V_0}{\partial R_0} \Delta R_0 \simeq \frac{V_0}{R_0} \simeq \frac{V_0}{\xi} \sim \frac{U}{\xi^2}
\end{align}
where for the last two approximate equalities we have used the fact that the size of the droplet $R_0\sim R_0^{\mathrm{max}}$ and  $V_0\xi \sim U$.

The SF-Mott transition can occur when either the size of the droplet is changed by $\Delta R_0 \sim 1$ or upon rearranging the density profile for $\Delta R_0 \ll 1$, depending on what happens first in the process of changing of the parameters of the Hamiltonian.
In the former case we have $\Delta V_0^{\mathrm{Mott-SF}} = \Delta V_0^{\mathrm{dr}}$.
In the latter case the SF-Mott transition corresponds to redistributing $\sim 1$ particle from the center of the droplet to its edges (see Fig.~\ref{suppl_fig:Thomas-Fermi}(b) for the change of $n(x)$ upon changing the parameters of the Hamiltonian):
\begin{align}
\frac{\partial n(0)}{\partial V_0} \Delta V_0 \simeq \frac{1}{N}
\end{align}
\begin{align}
\Delta V_0^{\mathrm{Mott-SF}} \simeq \frac{V_0 R_0}{N} \simeq  \frac{V_0 \xi}{N}  \sim \frac{U}{N}
\label{SupplEq:dV_Mott-SF}
\end{align}

The last approximate equality holds for droplets of characteristic maximal size $R_0 \sim R_{max}$ of the Thomas-Fermi regime. It is convenient to introduce the parameter
\begin{align}
\sigma = \frac{N}{\xi^2} \sim \frac{ \Delta V_0^{\mathrm{dr}} } { \Delta V_0^{\mathrm{Mott-SF}} }  
\label{SupplEq:steepness}
\end{align}
On the one hand, it gives an estimate for the number of SF- and Mott-droplet stripes that can fit within a droplet phase with fixed size.
On the other hand, it gives an estimate of the spatial derivative of the density of the droplet $\sigma \sim \partial n(x)/\partial x \sim (N/\xi)/\xi = N/\xi^2$. We thus name $\sigma$ the droplet \emph{steepness} parameter. According to it we classify the droplets as  \emph{steep} ($\sigma\gg1$) or \emph{flat}  ($\sigma\ll1$).

Note that in the thermodynamic limit for the droplet, $N,\xi\rightarrow\infty$, with $N/\xi=n_{0}=\mathrm{const}$, the steepness parameter $\sigma=n_{0}/\xi \rightarrow 0$, so only the flat droplets survive. In this regime eq. (\ref{SupplEq:dV_Mott-SF}) (except of the last proportionality specific to the Thomas-Fermi limit) still gives a good estimate for the characteristic width of SF/Mott droplet regions, from which we see that all the  phases with the droplet radius scaling as $R\sim\xi$ survive in the thermodynamic limit. At the same time, all the regions of the phase diagram corresponding to the droplets of finite fixed size $R$ shrink in the thermodynamic limit as $\Delta V_0^{\mathrm{Mott-SF}} \sim 1/N$ (see first equality in eq. (\ref{SupplEq:dV_Mott-SF})).

In the main text we study the only the flat droplet regime relevant to the thermodynamic limit, while the steep droplets might be present in finite systems also relevant for experiments and we discuss them in the next Supplementary Section.

\section{Finite-size regime of steep droplets}

\subsection{Monotonous interaction in 2D}

Performing the same analysis done for the flat-droplet regime in the main text,  we consider here the steep droplet regime, both for the monotonous interaction in 1D and for the sign-changing interaction in 2D.

\begin{figure}[!h] 
	\centering
	\includegraphics[width=1.0\linewidth]{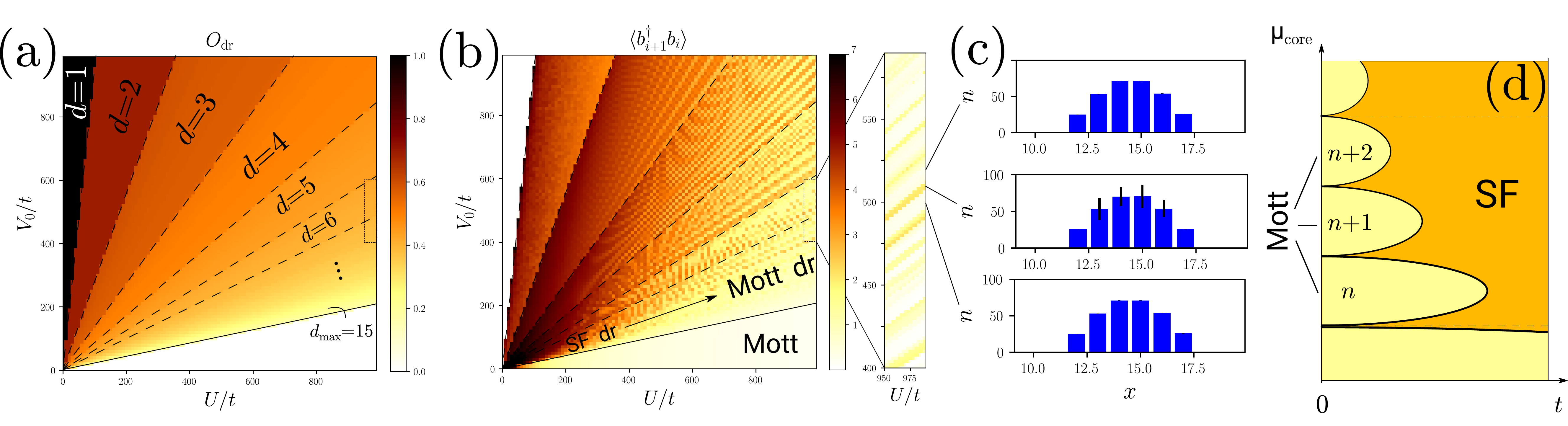}
	\caption{
		Steep droplet regime for monotonous interaction in 1D.
		Low-temperature phase diagrams in the coordinates $V_0$ vs $U$ for  $\sigma=N/\xi^2\approx33 \gg1$ ($L=30$, $N=300$, $\xi=3$). 
		(a) Droplet order parameter  $O_{\mathrm{dr}}$. 
		(b) Nearest-neighbor superfluid correlator $\langle b^{\dagger}_{i+1} b_i \rangle$. 
		(c) Spatial profiles of the droplets corresponding to the points in the phase diagrams Fig. (b). The errorbars show the standard deviation of density from its mean value -- they are stretched by factor of 100 for the visual clarity.
		(d) Schematic standard Bose-Hubbard phase diagram.
		Dashed lines in all figures are guides for the eye for the droplet-size change mesoscopic phase transitions, vanishing in the thermodynamic limit. Solid lines  show the transitions present also in the thermodynamic limit. SF phase in (a), (b) are in the region $U/t,V_0/t\lesssim 10$ (not seen at the given scale).
	}
	\label{SupplFig:steep_droplet_monotonous_1D}
\end{figure}

First we consider the steep droplet regime for the monotonous interaction in 1D.
For $U=0, V_0>0$ we are the single-site droplet phase. Upon increasing $U$ and decreasing $V_0$ we observe a sequence of first-order phase transitions between droplets of different ``diameters'' (Fig.~\ref{SupplFig:steep_droplet_monotonous_1D}(a)), up to the maximum diameter bounded either by the size of the system or the total number of particles $d_{\mathrm{max}}\simeq \min(L/2, N)=15$.
We observe that, as soon as $d$ exceeds $\min(\xi, N)$, the corresponding region of the phase diagram shrinks with increasing $d$, indicating the fact that $d=\min(\xi, N)$ is the characteristic diameter of the droplet.

For steep droplets, each region of the phase diagram corresponding to a droplet of a given size further subdivides into smaller regions (Fig.~\ref{fig:phase-diagram_monotonous}(b)). These finer transitions correspond to a redistribution of the droplet density profile while its diameter stays fixed. Between two stable configurations with almost non-fluctuating local particle densities (Mott states), which correspond to different discretizations of the optimal continuous droplet profile (upper and lower density profiles in Figs.~\ref{fig:phase-diagram_monotonous}(c)), there is always a superfluid region with local density fluctuations.

The steepness parameter $\sigma$ determines how many different local occupations of the core are compatible with a given droplet size (Fig.~\ref{fig:phase-diagram_monotonous}(b)), that is, how many Mott lobes fit into a given sector of fixed $d$ in  Fig.~\ref{fig:phase-diagram_monotonous}(a) (see the discussion in the previous Supplementary Section). 

\subsection{Sign-changing interaction in 2D}

\begin{figure}[!h] 
	\centering
	\includegraphics[width=0.6\linewidth]{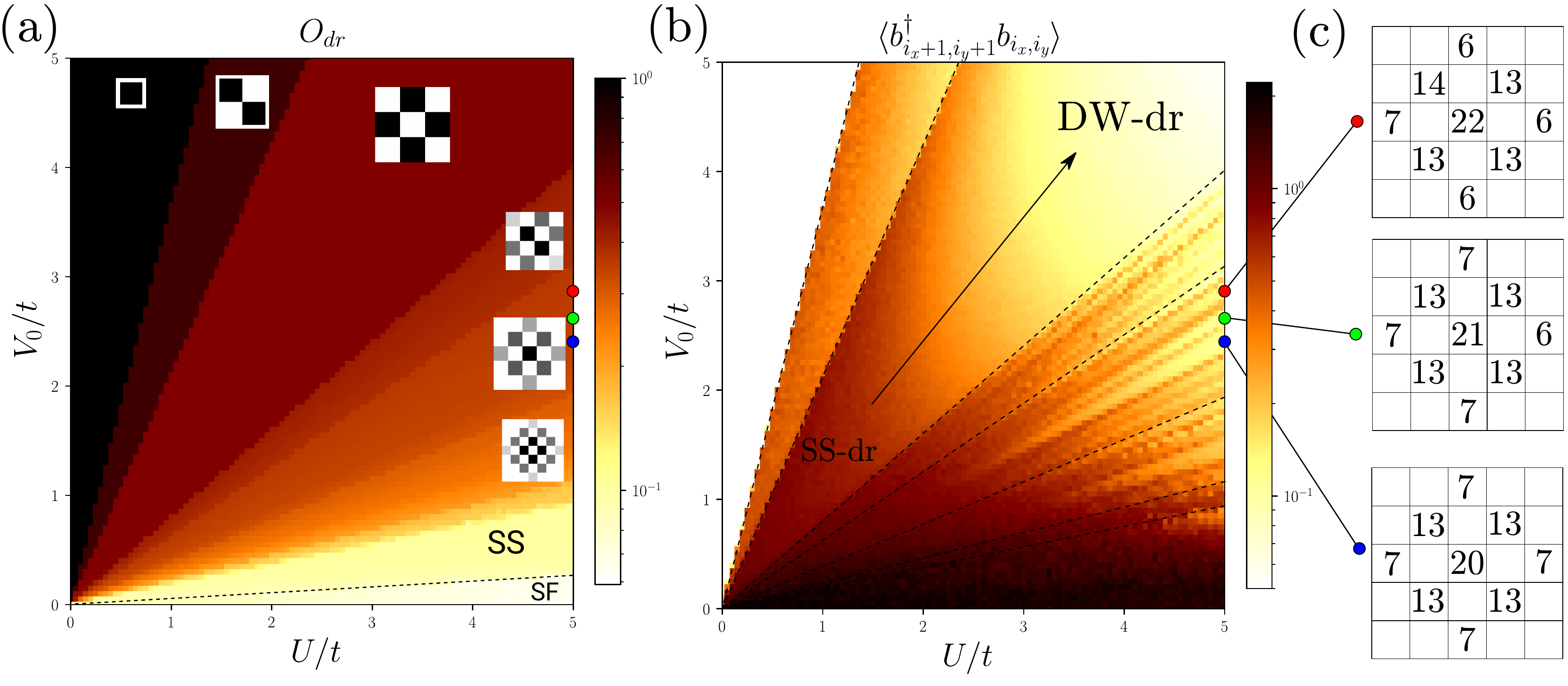}
	\caption{
		Steep droplet regime for sign-changing interaction in 2D.
		Low-temperature phase diagram in the coordinates $V_0$ vs $U$ for $\sigma=N/\xi^3=6.4$ ($L=10$, $N=100$, $\xi=2.5$). 
		(a) Droplet order parameter  $O_{\mathrm{dr}}$ (the insets show spatial structures of the droplets). 
		(b) Superfluid correlator $\langle b^{\dagger}_{i_x+1,i_y+1} b_{i_x,i_y} \rangle$. (c) Spatial structures of the droplets at the corresponding points of phase diagram; the numbers show the local occupation numbers in the Mott regime (for the unmarked sites $n_{\mathbf{i}}=0$).
	}
	\label{SupplFig:steep_droplet_sign_changing_2D}
\end{figure}

Now we turn to the steep droplet regime for the sign-changing interaction in 2D.
Like for the monotonous interaction, for each droplet phase we observe a structure of superfluid radial lines in the phase diagram, i.e. a non-trivial change in the superfluid correlator in the azimuthal direction (Fig.~\ref{SupplFig:steep_droplet_sign_changing_2D}(b)). It is due to the fact that different, one or another discretization of the density profile $n(\mathbf{r})$ becomes more favorable, giving rise to different stable Mott structures (like in Fig.~\ref{SupplFig:steep_droplet_sign_changing_2D}(c)). In the middle of the crossover between two such structures we observe an increase in the superfluid correlators.

\section{Derivation of the extended Bose-Hubbard model}

\subsection{Cavity-mediated interaction}

In this Supplementary Section we describe the setting we propose
for the experimental realization of the interaction we use in the main
text. Apart from slight variations in the laser-beam orientation, the setup we propose is the one implemented with ultracold bosons in a confocal cavity in Stanford \cite{Vaidya:2018,Guo:2019,Guo:2019A,Guo:2021}.

We consider an ensemble of atoms treated as two-level systems with level splitting $\omega_a$, placed in a multimode optical cavity. The atomic transition is off-resonantly driven by either one or two laser beams oriented in the direction transverse to the cavity axis. Each beam has a frequency $\omega_L$ and the total laser field is indicated by $\Omega(\mathbf{r})$.
The cavity possesses (nearly) degenerate modes with frequencies $\omega_{c\alpha}$ (where $\alpha$ is the mode index).
The coupling between the atomic transition and a given cavity mode is given by $g_{\alpha} (\mathbf{r}) = g_{0} \Xi_{\alpha}(\mathbf{r})$, where $\Xi_{\alpha}(\mathbf{r})$ is the normalized mode function of the cavity mode.
We work in the dispersive regime, where the detuning $\Delta_{c\alpha} = \omega_{c\alpha} - \omega_L$ between the laser and the cavity degenerate family is much smaller than the detuning $\Delta_{a} = \omega_a - \omega_{L}$ between the laser and the atomic transition. In this case, we can adiabatically eliminate the atomic excited state and work only with itinerant atoms all the time in the electronic ground state. The optical potential that the atom experience is induced by two-photon transitions \cite{Piazza:2020}. The latter can either involve two laser-photons and give rise to a static spatially-dependent Stark shift 
	\begin{align}
	V_L(\mathbf{r})= -\frac{1}{\Delta_a} |\Omega(\mathbf{r})|^2 \label{eq:ExternalPotential}
	\end{align}
	or involve one photon from the laser and one from the cavity, which induces a dynamical optical potential whose spatial dependence results from the interference between the laser and the cavity field. We neglect the dynamical potential resulting purely from cavity photons as it can be suppressed by increasing the detuning $\Delta_{a}$ \cite{Vaidya:2018}. Assuming now that the cavity detuning $\Delta_{c\alpha}$ is larger than the dispersive coupling and the atomic motional scales we can further eliminate the cavity field adiabatically and obtain the following cavity-mediated interaction \cite{Vaidya:2018,Guo:2019,Guo:2019A,Piazza:2020}
	\begin{align}
	V(\mathbf{r},\mathbf{r}') =- \frac{g_0^2}{\Delta_a^2} \Omega^*(\mathbf{r}) \Omega(\mathbf{r}') \sum_{\alpha}  \frac{\Xi^*_{\alpha}(\mathbf{r}) \Xi_{\alpha}(\mathbf{r}')}{\Delta_{c\alpha}} \equiv - \frac{g_0^2}{\Delta_a^2} \Omega^*(\mathbf{r}) \Omega(\mathbf{r}') v(\mathbf{r},\mathbf{r}') \label{eq:InteractionPotential}
	\end{align}
	In order to have access to superradiant phase transitions and the corresponding spatial ordering, the laser must be red-detuned from the cavity degenerate family $\Delta_{c\alpha}>0$. In this case the interaction is attractive at small distances $\mathbf{r}'\rightarrow \mathbf{r}$ i.e. $V(\mathbf{r},\mathbf{r})<0$, which creates the possibility for droplet-formation.

In the main text, we consider the generic ansatz Eq.(2) for the interaction potential, decaying with the characteristic length $\xi$. Such translationally-invariant potentials can be directly created by concentric optical cavities, which are however experimentally not available in combination with an ultracold atomic cloud. Confocal cavities are instead already available in Standford and have been shown to mediated tunable range interactions \cite{Vaidya:2018}. Differently from our ansatz, the interaction in that case contains an additional non-translationally-invariant contribution of the type $\cos(\mathbf{r} \mathbf{r}')$. However, using a second laser near resonant with a second nearby degenerate family of the same parity, this term can be eliminated, as has been experimentally demonstrated in \cite{Guo:2021}. Finally, there is a further contribution not included in our model, which comes from the interaction of the atoms with their images on the other side of the cavity axis. This contribution can be also eliminated by placing the cloud sufficiently away (at distances $\gg \xi$) from the cavity axis. This has been also experimentally demonstrated in \cite{Guo:2021}.

\subsection{Lattice geometry}
\label{app:geometry}

We consider the situation where the atomic cloud is trapped within a deep optical lattice
\begin{align}
V_{\mathrm{ext}}(\mathbf{r}) &= -\frac{1}{\Delta_{\mathrm{ext}} } |\Omega_{\mathrm{ext}} (\mathbf{r})|^2 \label{eq:ExternalPotential}
\end{align}
formed by an additional set of laser beams, building the field $\Omega_{\mathrm{ext}} (\mathbf{r})$, with a detuning $\Delta_{\mathrm{ext}} = \omega_a - \omega_{\mathrm{ext}}$ from the atomic transition which is sufficiently different from $\Delta_a$ to avoid interference with the lasers involved in the interaction. Such a condition is achieved in state of the art experiments with ultracold bosons in a cavity \cite{Landig:2016}.
In what follows, we will assume that $|V_{\mathrm{ext}}(\mathbf{r})| \gg |V_L(\mathbf{r})|$ and hence neglect the latter.

\vspace{0.5cm}
\subsubsection{One-dimensional setting}

We choose the $z$-axis to be parallel to the cavity axis. We will use one laser beam for inducing the cavity-mediated interactions and three lasers for creating
the external optical lattice that confines the gas to well-separated 1D chains, parallel to the $x$-axis (see Fig.~\ref{fig:geometry} in the main text). The external optical potential reads
\begin{align}
V_{\mathrm{ext}}(\mathbf{r}) = -\frac{\Omega^{x}_{\mathrm{ext}} {}^2}{\Delta_{\mathrm{ext}}}\cos^2 (k_{\mathrm{ext}}x)  -\frac{\Omega^{yz}_{\mathrm{ext}} {}^2}{\Delta_{\mathrm{ext}}}(\cos^2 (k_{\mathrm{ext}}y)  + \cos^2 (k_{\mathrm{ext}}z) ) 
\end{align}
If $\Omega^{yz}_{\mathrm{ext}} \gg \Omega^{x}_{\mathrm{ext}}$, the two latter terms confine the gas to the tubes $y=z=0$.
\\

a) \textit{Sign-changing interaction}. The interaction-inducing laser is oriented along the $x$-axis: $\Omega(\mathbf{r})=\Omega\cos(kx)$. The resulting interaction potential reads
\begin{align}
V(\mathbf{r}, \mathbf{r}') = -\frac{g_0^2\Omega^2}{\Delta_a^2}\cos(kx) \cos(kx') v(\mathbf{r}, \mathbf{r}')
\end{align}
Let $k_{\mathrm{ext}}\approx k=2\pi/\lambda$.
For $\Delta_{\mathrm{ext}}>0$ the external potential confines the gas near points $kx=\pi n \Leftrightarrow x=n \lambda/2$  ($n\in \mathbb{Z}$). 
Near such points $\mathbf{r}_n=(n\lambda/2,0,0)$ we have $V(\mathbf{r}_n, \mathbf{r}_{n'}) \sim (-1)^n (-1)^{n'} v(\mathbf{r}_n-\mathbf{r}_{n'})=(-1)^{n-n'} v(\mathbf{r}_n-\mathbf{r}_{n'})$, which is the desired 1D sign-changing interaction.

b) \textit{Monotonous interaction}. Now we orient the interaction-inducing laser along the $y$-axis. Since we consider a deep trapping lattice such the atoms are strongly confined along $y,y'=0$, the sign-changing part of the interaction is approximately constant: $\cos(ky) \cos(ky')=1$ and we obtain the 1D monotonous attractive interaction considered in the main text:
\begin{align}
V(\mathbf{r}, \mathbf{r}') = -\frac{g_0^2 \Omega^2}{\Delta_a^2} v(\mathbf{r}, \mathbf{r}')
\end{align}

\subsubsection{2D setting}

Here we use two interaction-inducing laser beams $\mathbf{\Omega}_1(\mathbf{r}) =\mathbf{e}_1\Omega \cos kx$,  $\mathbf{\Omega}_2(\mathbf{r}) =\mathbf{e}_2\Omega \cos ky$, while the trapping lattice is formed by the following beams: $\mathbf{\Omega}_{\mathrm{ext},1}(\mathbf{r}) = \tilde{\mathbf{e}}_1\Omega^{xy}_{\mathrm{ext}} \cos kx$,  $\Omega_{\mathrm{ext},2}(\mathbf{r}) = \tilde{\mathbf{e}}_2\Omega^{xy}_{\mathrm{ext}} \cos ky$, $\Omega_{\mathrm{ext},3}(\mathbf{r}) = \tilde{\mathbf{e}}_3 \Omega^{z}_{\mathrm{ext}} \cos kz$. The optical lattice and the interaction potentials thus read
\begin{align}
V_{\mathrm{ext}}(\mathbf{r}) &= -\frac{1}{\Delta_a}(\Omega^{xy}_{\mathrm{ext}} \tilde{\mathbf{e}}_1\cos (k_{\mathrm{ext}} x) + \Omega^{xy}_{\mathrm{ext}} \tilde{\mathbf{e}}_2\cos (k_{\mathrm{ext}} y) + \Omega^{z}_{\mathrm{ext}} \tilde{\mathbf{e}}_3\cos (k_{\mathrm{ext}} z))^2\\
V(\mathbf{r},\mathbf{r}') &= -\frac{g_0^2 \Omega^2}{\Delta_a^2} (\mathbf{e}_1\cos (kx) + \mathbf{e}_2 \cos (ky)) (\mathbf{e}_1\cos (kx') + \mathbf{e}_2 \cos (ky')) v(\mathbf{r},\mathbf{r}')
\end{align}

\begin{figure}[!tbh] 
	\centering
	\includegraphics[width=0.7\linewidth]{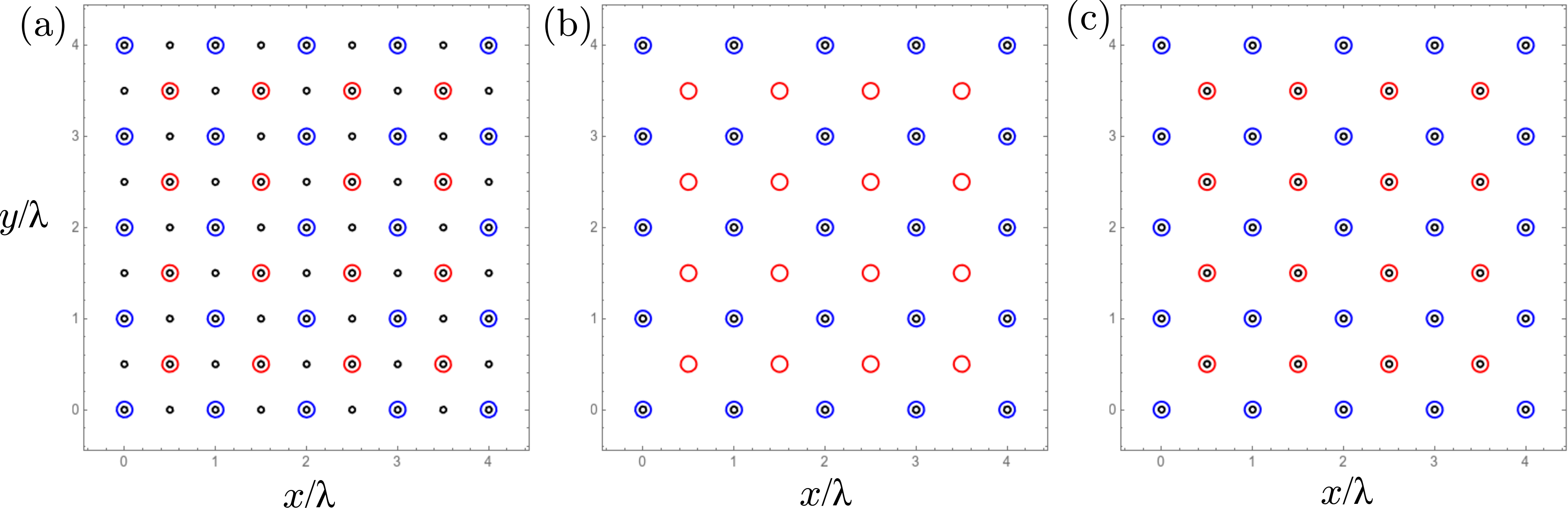}
	\caption{
		Periodicities of the interaction potential (red and blue circles) and of external optical lattice (black circles) in the 2D setting for (a) the $Z_4$ interaction, (b) the monotonous interaction, and (c) the sign-changing interaction considered in the main text.
	}
	\label{suppl_fig:geometry_2D}
\end{figure}

We chose the polarizations of the interaction-inducing lasers to be $\mathbf{e}_1 = \mathbf{e}_y \perp \mathbf{e}_2 = \mathbf{e}_x$, so they can scatter into the cavity.
Let us also first assume that for the optical lattice lasers we have $\tilde{\mathbf{e}}_1 \perp \tilde{\mathbf{e}}_2 \perp \tilde{\mathbf{e}}_3$ and $\Omega^{z}_{\mathrm{ext}}  \gg \Omega^{xy}_{\mathrm{ext}}$. We obtain
\begin{align}
V_{\mathrm{ext}}(x,y,0) &= -\frac{ \Omega^{xy}_{\mathrm{ext}} {}^2}{\Delta_a}(\cos^2 (k_{\mathrm{ext}}x) + \cos^2 (k_{\mathrm{ext}} y)) - \frac{ \Omega^{z}_{\mathrm{ext}} {}^2}{\Delta_a} \cos^2 (k_{\mathrm{ext}}z) \\
V(x,y,0; x',y',0) &= -\frac{g_0^2 \Omega^2}{\Delta_a^2} (\cos (kx)\cos (kx') + \cos (ky) \cos (ky')) v(\mathbf{r},\mathbf{r}')
\end{align}
so the gas is restricted to the pancake geometry ($z=0$ plane) and the minima of the external potential are $\mathbf{r}_{n,m} =  (n \lambda_{\mathrm{ext}}/2, m \lambda_{\mathrm{ext}}/2, 0)$ (where $n,m\in \mathbb{Z}$). 

a) \emph{$Z_4$ interaction}. We choose $k_{\mathrm{ext}} \approx k$. $V(\mathbf{r}_{n,m}, \mathbf{r}_{n',m'})\sim - ((-1)^n(-1)^{n'}+(-1)^m(-1)^{m'})= -((-1)^{n-n'} +(-1)^{m-m'})$, so the interaction potential is translation invariant on the set of all points of the optical lattice as well as all its subsets. We show the periodicity of this interaction in Fig.~\ref{suppl_fig:geometry_2D}(a): if both particles sit on the same checkerboard sublattice their interaction energy is either positive or negative; if they sit on the different sublattices, their interaction energy is zero. We denote this case as $Z_4$ interaction.

b) \emph{Monotonous interaction}. If we choose instead the frequencies of the external optical lattice lasers and the interaction-inducing lasers in such a way that $\omega_{\mathrm{ext}} = 2\omega$ (so, $k_{\mathrm{ext}} = k/2$), then the period of the external optical lattice would be doubled and we obtain a translation-invariant monotonous interaction (Fig.~\ref{suppl_fig:geometry_2D}(b)).

c) \emph{Sign-changing interaction}. This is obtained by choosing $\tilde{\mathbf{e}}_1 || \tilde{\mathbf{e}}_2 \perp \tilde{\mathbf{e}}_3$ and  $k_{\mathrm{ext}} \approx k$. In this case thus the lattice lasers 1 and 2 interfere, so that $V_{\mathrm{ext}}(x,y,0)= -\frac{\Omega^{xy}_{\mathrm{ext}}{}^2}{\Delta_a}( \cos (k_{\mathrm{ext}}x) + \cos (k_{\mathrm{ext}}y))^2$ and we obtain the lattice configuration shown in Fig.~\ref{suppl_fig:geometry_2D}(c). It corresponds to the 2D sign-changing interaction considered in the main text. Note that the optical lattice axes $\tilde{x}\tilde{y}$ are $45^{\circ}$ rotated with respect to the $xy$ axes.

\end{document}